\documentclass[pra,superscriptaddress,nofootinbib,twocolumn]{revtex4-1}
\usepackage{amsmath,graphicx,epsfig,color,amsfonts,amssymb,bbm}
\usepackage{amsthm}
\usepackage{mathtools}
\usepackage{hyperref}

\newcommand{\etal}{{\em et al. }}

\newcommand{\ket}[1]{|#1\rangle}

\newcommand{\id}{\mathbbm{1}}
\renewcommand{\S}{\mathcal{S}}
\newcommand{\vecP}{\vec{P}}
\newcommand{\bfx}{\mathbf{x}}
\newcommand{\NS}{\mathcal{NS}}
\newcommand{\B}{\mathcal  B}
\newcommand{\R}{\mathcal{R}}

\begin{document}

\title{Anonymous Quantum Nonlocality}

\author{Yeong-Cherng~Liang}
\affiliation{Institute for Theoretical Physics, ETH Zurich, 8093 Zurich, Switzerland.}
\author{Florian John Curchod}
\affiliation{ICFO--Institut de Ci\`encies Fot\`oniques, 08860 Castelldefels (Barcelona), Spain.}
\affiliation{Group of Applied Physics, University of Geneva, CH-1211 Geneva 4, Switzerland.}
\author{Joseph Bowles}
\affiliation{Department of Theoretical Physics, University of Geneva, 1211 Geneva, Switzerland.}
\author{Nicolas Gisin}
\affiliation{Group of Applied Physics, University of Geneva, CH-1211 Geneva 4, Switzerland.}

\date{\today}

\begin{abstract}
We show that for all $n\ge3$, an example of an $n$-partite quantum correlation that is {\em not} genuinely multipartite nonlocal but rather  exhibiting {\em anonymous nonlocality}, that is, nonlocal but biseparable with respect to {\em all} bipartitions, can be obtained by locally measuring  the $n$-partite Greenberger-Horne-Zeilinger (GHZ) state. This  anonymity is a manifestation of the impossibility to attribute unambiguously the underlying multipartite nonlocality to  {\em any definite subset(s)} of the parties, even though the correlation can indeed be produced by nonlocal collaboration involving only such subsets.
An explicit biseparable decomposition of these correlations is provided  for {\em any} partitioning of the $n$ parties into two groups.
Two possible applications of these anonymous GHZ correlations in the {\em device-independent} setting are discussed: multipartite secret sharing between any two groups of parties and bipartite quantum key distribution that is robust against nearly arbitrary leakage of information.
\end{abstract}

\maketitle

Quantum correlations that violate a Bell inequality~\cite{Bell1964}, a constraint first derived in the studies of local-hidden-variable-theories, were initially perceived only as a counterintuitive feature with no classical analog. With the discovery of quantum information science, these intriguing correlations have taken the new role as a resource. For instance,  in  nonlocal games~\cite{Cleve2004}, the presence of Bell-inequality-violating (hereafter referred as nonlocal) correlation signifies the usage of strategies that cannot be achieved using only shared randomness. They are also an indispensable resource in quantum information  and communication tasks such as the reduction of communication complexity~\cite{CommComplexity}, the distribution of secret keys using untrusted devices~\cite{DIQKD,Acin:NJP:2006}, as well as the certification and expansion of randomness~\cite{Randomness} etc. (see~\cite{Brunner:RMP} for a  review).
 
Thus far, prior studies of quantum nonlocality have focussed predominantly on the bipartite setup. However, as with quantum entanglement~\cite{Entanglement,Multipartite}, 
 correlations between measurement outcomes can exhibit a much richer structure in the multipartite setup. Consider a multipartite Bell-type experiment with the $i$-th party's choice of measurement setting (input)  denoted by $x_i=0,1$ and the corresponding outcome (output) by $a_i=\pm1$. Already in the tripartite setting~\cite{Svetlichny}, quantum mechanics allow for correlations --- a collection of joint conditional probability distributions $\vecP=\{P(\vec{a}|\vec{x})\}=\{P(a_1a_2a_3|x_1x_2x_3)\}$  ---  that cannot be reproduced even  when a subset of the parties are allowed to share some nonlocal resource $\R$~\cite{Gallego:PRL:070401,Bancal:PRA:014102}.\footnote{Throughout, we  focus  on nonlocal resources $\R$ that respect the non-signaling conditions~\cite{NS,Barrett:PRA:2005} which dictate, e.g., that each marginal of distribution of $P^\R_i(a_ja_k|x_jx_k)$  can be defined independent of the input of the other party.}
Such {\em genuinely tripartite nonlocal} correlations are, by definition, those that {\em cannot} be written in the so-called {\em biseparable} form:
\begin{equation}\label{Eq:Bisep}
\begin{split}
	P(\vec{a}|\vec{x})&\neq\sum_\nu p_\nu\, P_\nu(a_1|x_1)P^\R_\nu(a_2a_3|x_2x_3)\\
			    	&+\sum_\mu p_\mu\, P_\mu(a_2|x_2) P^\R_\mu(a_1a_3|x_1x_3) \\
				&+\sum_\lambda p_\lambda\, P_\lambda(a_3|x_3) P^\R_\lambda(a_1a_2|x_1x_2) 
\end{split}
\end{equation}
where $\sum_{i\in\{\lambda,\mu,\nu\}} p_i =1$, $p_{i} \ge 0$ for all $i \in \{\lambda, \mu, \nu \}$ and $P^\R_i(a_ja_k|x_jx_k)$ is {\em any} 2-partite correlation  allowed by the resource $\R$~\cite{Gallego:PRL:070401,Bancal:PRA:014102}.
In a Bell-type experiment, the presence of genuine multipartite nonlocality~\cite{GMNL,Bancal:PRL:2009,Aolita:PRL:2012,Bancal:JPA:2012,QChen:PRL:2014} is a manifestation of genuine multipartite entanglement~\cite{Entanglement}, thus facilitates the detection of the latter in a {\em device-independent} manner, i.e., without relying on any assumption about the measurements being performed nor the dimension of the underlying Hilbert space.\footnote{It is also possible to detect genuine multipartite entanglement in a device-independent manner without the detection of genuine multipartite nonlocality. See~\cite{DIEW,Bancal:JPA:2012}.} In contrast, correlations that are biseparable, cf. Eq.~\eqref{Eq:Bisep}, receive almost no attention. Apart from being a tool in the derivation of Bell-type inequalities for genuine multipartite nonlocality, is this kind of correlations interesting in its own right? Here, we answer this question affirmatively via the phenomenon of {\em anonymous nonlocality} (ANL), an intriguing feature that is only present in  biseparable correlations. We will  also provide evidence showing that ANL can be a powerful resource, allowing one to design device-independent quantum cryptographic protocols that can guard against a particular kind of attack by any post-quantum, but non-signaling  adversary.

{\em Biseparable correlations and anonymous nonlocality.}-- To appreciate the peculiarity manifested by ANL, let us start by considering the simplest, tripartite scenario. Clearly, among the subsets of correlations that can be decomposed in the form of the right-hand-side of Eq.~\eqref{Eq:Bisep} are those that satisfy:
\begin{subequations}\label{Eq:DN1}
\begin{align}
	P(\vec{a}|\vec{x}) &=\sum_\nu p_\nu P_\nu(a_1|x_1) \, P^\R_\nu(a_2a_3|x_2x_3) ,\label{Eq:Bisep:BC-A}\\
	 &=\sum_\mu p_\mu\, P_\mu(a_2|x_2) P^\R_\mu(a_1a_3|x_1x_3) ,\label{Eq:Bisep:AC-B}\\
					&=\sum_\lambda p_\lambda\, P_\lambda(a_3|x_3) P^\R_\lambda(a_1a_2|x_1x_2) ,\label{Eq:Bisep:AB-C}
\end{align}
where $p_\nu, p_{\mu}, p_\lambda \ge 0$ for all $\nu$, $\mu$ and $\lambda$, but in contrast with Eq.~\eqref{Eq:Bisep}, we now have $\sum_\nu p_\nu =\sum_\mu p_\mu = \sum_\lambda p_\lambda =1$. Eqs.~\eqref{Eq:Bisep:BC-A}--\eqref{Eq:Bisep:AB-C} imply that the correlation can be produced {\em without} having any nonlocal collaboration between the isolated  party and the remaining two parties (as a group). Naively, one may thus  expect that all correlations satisfying these equations must also be Bell-local (henceforth abbreviated as local). However, there exist~\cite{Vertesi:PRL:030403} quantum correlations that satisfy Eqs.~\eqref{Eq:Bisep:BC-A}--\eqref{Eq:Bisep:AB-C} as well as:
\begin{equation}
	P(\vec{a}|\vec{x}) \neq\sum_\theta p_\theta\, P_\theta(a_1|x_1) P_\theta(a_2|x_2) P_\theta(a_3|x_3),\label{Eq:Nonlocal}
\end{equation}
\end{subequations}
for {\em any} conditional distributions $P_\theta(a_i|x_i)$ and {\em any} normalized weights $p_\theta$. In other words, $\vecP$ satisfying Eq.~\eqref{Eq:DN1} is nonlocal but this nonlocality is (i) not genuinely tripartite (it is biseparable) (ii) not attributable to any of the two-partite marginals [Eqs.~\eqref{Eq:Bisep:BC-A}--\eqref{Eq:Bisep:AB-C} imply that all marginals are local] and (iii) not attributable to any bipartition of the three parties. The nonlocality present in {\em any} correlations satisfying Eq.~\eqref{Eq:DN1} is thus in some sense nowhere to be found!

We now provide a very simple example of correlation satisfying Eq.~\eqref{Eq:DN1}, and more generally the property of being (1) nonlocal and (2) biseparable with respect to all bipartitions in an arbitrary $n$-partite scenario. Consider the $n$-partite Greenberger-Horne-Zeilinger (GHZ) state~\cite{GHZ} $\ket{\rm GHZ_n} = \frac{1}{\sqrt{2}}  (\ket{0}^{\otimes n}+\ket{1}^{\otimes n})$ and the local measurement of $\sigma_x$ and $\sigma_y$. The resulting correlation is
\begin{equation}\label{Eq:GHZ:Correlations}
	P(\vec{a}|\vec{x})=P^n_{\rm GHZ}(\vec{a}|\vec{x})=\frac{1}{2^n}\left[1+\cos\left(\bfx\frac{\pi}{2} \right)\prod_{i=1}^n a_i\right],
\end{equation} 
where $\bfx=\sum_i x_i$, and we have identified $x_i=0\, (1)$ as the $\sigma_x (\sigma_y)$ measurement (see, eg. Eq.~(23) of ~\cite{Wallman:PRA:2011}).
In Appendix~\ref{App:NS-Bisep}, we show that {\em for all} $n\ge 3$, $n$-partite correlations of the form of Eq.~\eqref{Eq:GHZ:Correlations} admit a biseparable decomposition with respect to {\em any} partitioning of the $n$ parties into two groups. Specifically for $n=3$, this decomposition, cf.  Eq.~\eqref{Eq:Bisep:BC-A}, involves $p_\nu = \tfrac{1}{4}$  for all  $\nu$, $P_\nu(a_1|x_1)=0,1$ and $P^\R_\nu(a_2a_3|x_2x_3)$ is the correlation associated with the so-called Popescu-Rohrlich (PR) box~\cite{NS} --- a hypothetical, stronger-than-quantum, but non-signaling resource.\footnote{In the {\em tripartite} scenario, the biseparability of the  GHZ correlation was also discovered independently in~\cite{Pironio:JPA:2011} (see also~\cite{Bisep}).} To see that these correlations are nonlocal, it suffices to note that Eq.~\eqref{Eq:GHZ:Correlations} violates the Mermin-Bell inequality~\cite{MABK1,MABK2} (even maximally~\cite{Werner:PRA:2000}  for all odd $n\ge 3$). See Appendix~\ref{App:BIV}.

\begin{figure}[h!]
\scalebox{0.5}{\includegraphics{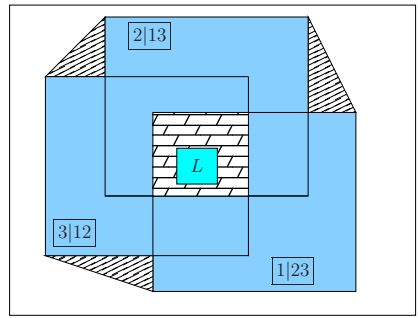}}
\caption{\label{Fig:Polytopes}
(Color online) Schematic representation of the various sets of tripartite correlations. Correlations biseparable with respect to party $i$ in one group and parties $j$ and $k$ in the other  lie in  the (light blue) rectangle labeled by ``$i|jk$". The convex hull of the three biseparable sets ``$i|jk$" where $i,j,k\in\{1,2,3\}$ is represented by the filled convex region and gives correlations decomposable as the right-hand-side of Eq.~\eqref{Eq:Bisep}. The blank region between the outermost box and the filled convex region represents correlations that are genuinely tripartite nonlocal. Intersection of the three biseparable subsets ``$i|jk$" gives correlations satisfying Eqs.~\eqref{Eq:Bisep:BC-A}--\eqref{Eq:Bisep:AB-C}; its subset  featuring ANL is the tiled region while local correlations lie in the  (cyan) rectangle $L$. Hatched regions represent biseparable correlations where classical mixture of different bipartitions is necessary for their production.  }
\end{figure}

Consider now an alternative way to understand the nonlocality associated with Eq.~\eqref{Eq:DN1}. Operationally, Eq.~\eqref{Eq:Bisep:AB-C} implies that $\vecP$ can be produced by, e.g.,  party 1  signaling classically to party 2, and all parties responding according to the information  that they received and some predefined strategy $\lambda$. By symmetry of Eqs.~\eqref{Eq:Bisep:BC-A}--\eqref{Eq:Bisep:AB-C}, the same can be achieved by having only nonlocal collaboration between any two out of the three parties. Thus, while the correlation can be produced by having only a {\em definite subset} of parties collaborating nonlocally, the identity of these nonlocally collaborating parties is {\em anonymous} to an outsider who only has access to $\vecP$. Indeed, even if an outsider is given the promise that a {\em fixed} subset of the parties have collaborated nonlocally, it is impossible for him to tell if, say,  party 1 and 2 have collaborated nonlocally  in generating $\vecP$. Importantly, the anonymity present in these correlations differs from the case where a {\em classical mixture} of the different bipartitions is necessary, cf. Fig.~\ref{Fig:Polytopes} (see~\cite{NS22,Curchod:IP} for examples of such classical anonymity). In this latter case, it is indeed possible to identify the parties that {\em must have} collaborated nonlocally, even though this identification is generally not possible at any single run of the experiment.

\begin{figure}[h!]
\scalebox{0.7}{\includegraphics{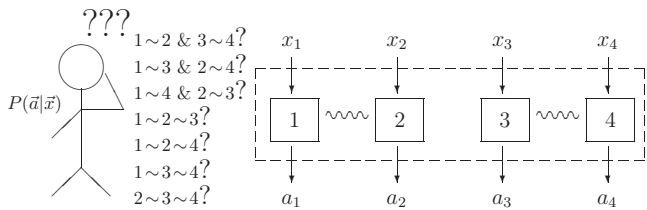}}
\caption{\label{Fig:ANL}
ANL in the 4-partite scenario. Each participating party is abstractly represented by a box labeled by the party number. The correlations were produced by having parties 1 \& 2, as well as  3 \& 4 collaborated  nonlocally (symbolized by ``$\sim$"). To an outsider who only has access to $\vec{a}$ and $\vec{x}$, even if one is given the promise that the correlations were produced by the four parties separated into two fixed groups, it is impossible to tell which actual partitioning of the parties generated these correlations. }
\end{figure}

As remarked above, for all $n\ge3$, the GHZ correlations of Eq.~\eqref{Eq:GHZ:Correlations} are nonlocal but  can nevertheless be produced by splitting the parties into {\em any} two groups, and disallowing any nonlocal collaboration between these groups. Thus, the anonymity present in these correlations is even more striking in the $n>3$ scenarios: not only are the groups of parties sharing $\R$ unidentifiable in an unambiguous manner, even the {\em size} of the groups are also not identifiable (see Fig.~\ref{Fig:ANL}). For example, when $n=4$, the  correlation satisfy:
\begin{subequations}\label{Eq:DN:4-partite}
\begin{align}
P(\vec{a}|\vec{x}) &= \sum_{\lambda_1} q_{\lambda_1} P_{\lambda_1}(a_1|x_1)P^\R_{\lambda_1}(a_2a_3a_4|x_2x_3x_4),\\
&= \sum_{\lambda_2} q_{\lambda_2} P_{\lambda_2}(a_2|x_2)P^\R_{\lambda_2}(a_1a_3a_4|x_1x_3x_4),\\
&=\cdots\,,\\
&= \sum_{\mu_3} q_{\mu_3} P^\R_{\mu_3}(a_1a_4|x_1x_4)P^\R_{\mu_3}(a_2a_3|x_2x_3),\\
P(\vec{a}|\vec{x}) &\neq \sum_{\theta} q_{\theta} \prod_{i=1}^4 P_{\theta}(a_i|x_i),
\end{align}
\end{subequations}
where $\sum_{\lambda_i} p_{\lambda_i} =\sum_{\mu_j} p_{\mu_j}=  1$, $p_{i} \ge 0$ for all $i,j$, and ``$\cdots$" indicates other possible biseparable decompositions that have been omitted. From Eq.~\eqref{Eq:DN:4-partite}, we see that the 4-partite GHZ correlation could have been produced by having {\em any} three parties collaborating nonlocally, or {\em any} two groups of two parties  collaborating nonlocally within each group. From the correlation itself, it is simply impossible to distinguish these possibilities apart  (Fig.~\ref{Fig:ANL}). 

Let us now briefly comment on the relationship between ANL and multipartite entanglement. Clearly, one expects that there must also be features  analogous to ANL in the studies of multipartite entanglement. Indeed, the first of such examples dates back to the three-qubit bound entangled~\cite{Horodecki:BE}  SHIFT state~\cite{Bennett:PRL:99} where its entanglement was dubbed {\em delocalized}~\cite{DiVincenzo} since it is separable with respect to all bipartitions, yet not fully separable. A more recent example~\cite{Vertesi:PRL:030403} involves a three-qubit bound entangled state which  even violates a Bell inequality, thus giving also an example of anonymous quantum correlation.
An important difference between their example and the tripartite case of our GHZ example  is that their correlation can be produced by a biseparable tripartite entangled state whereas ours necessarily requires a genuinely tripartite entangled state. More generally, for all odd $n\ge 3$, we show in Appendix~\ref{App:Quantum:BS} that the correlations of Eq.~\eqref{Eq:GHZ:Correlations} can only be produced by genuinely $n$-partite entangled state. Our examples thus show that the generation of ANL does not require delocalized entanglement.

{\em Perfect correlations with uniform marginals.}-- 
From Eq.~\eqref{Eq:GHZ:Correlations}, we see that whenever an odd number of parties measure in the $\sigma_y$ basis, the product of outcomes $\prod_i a_i$ gives $\pm1$ with equal probability, otherwise it is either perfectly correlated or perfectly anti-correlated. Moreover, it follows from Eq.~\eqref{Eq:GHZ:Correlations} that all marginal distributions of these correlations are uniformly random.
Next, we present two quantum cryptography protocols that exploit these strong but anonymous correlations.

{\em Application I: multipartite secret sharing (MSS).}-- 
Imagine that $n$ parties wanted to share a secret message between {\em any} two complementary subgroups as they desire, i.e., between any subgroup of $k$ parties ($k\le n-1$) and the subgroup formed by the remaining parties. Suppose moreover that the shared secret is to be recovered by these subgroups {\em only when all parties} within each group  collaborate (so that it is unnecessary to trust all parties within each group). A possibility to achieve this consists in: (i) the $n$ parties share (many copies of) $\ket{\rm GHZ_n}$, (ii) each party randomly measures either the $\sigma_x$ or the $\sigma_y$ observable, (iii) the $n$ parties are {\em randomly} separated into two groups and all parties assigned to the same group collaborate to compare their inputs and outputs, (iv) both {\em groups} announce their sum of inputs, (v) parties in the same group compute the product of their measurement outcome and deduce, using Eq.~\eqref{Eq:GHZ:Correlations},  the shared secret bit upon learning the sum of inputs of the other group, (vi)  parties in one group use the shared secret keys to encrypt the message and send it to the other. 

In the device-independent setting, security analysis is carried out by treating each physical subsystem together with their measurement device as a black box; conclusions are drawn directly from the measurement statistics. Indeed, the above protocol does not rely on the assumption of a GHZ state nor the particular measurements being performed,  but rather the strong correlation present in Eq.~\eqref{Eq:GHZ:Correlations} --- for the right combination of inputs, the product of outputs are perfectly (anti) correlated.\footnote{This happens in half the cases. In the other cases, the correlation is useless for key generation.}  Thus, the protocol essentially  works by first distributing the correlated  data needed to establish the secret keys, and performing the secret sharing~\cite{SS} between any two complementary subgroups of the $n$ participating parties as they deem fit. Since the product of outcomes for each group is uniformly random, the protocol is secure against cheating by any dishonest parties within the group; no one can retrieve the shared key without collaborating with everyone else within the same group. What about eavesdropping by an external, post-quantum but non-signaling adversary Eve?

Since the GHZ correlations of Eq.~\eqref{Eq:GHZ:Correlations} are biseparable, a naive attack by Eve may consist in preparing for the $n$ parties the biseparable, non-signaling boxes that reproduces exactly Eq.~\eqref{Eq:GHZ:Correlations}. For instance, in the tripartite case, in accordance with the biseparable decomposition, she would prepare with equal probability 4 different versions of a deterministic box for one of the parties, and correspondingly 4 different versions of a PR box for the remaining two parties. If the decomposition that she chooses matches exactly  the way the parties are separated into two groups, then after step (iv), she learns exactly the key and hence the message shared by these parties.\footnote{In this case, the product of outcomes for each group is a deterministic function (of the sum of inputs)  known to Eve. The secret sharing protocol of Hillery {\em et al.}~\cite{Hillery:PRA:1999} is thus insecure against this kind of attack by a non-signaling adversary.} 
However, as the grouping is decided only after the measurement phase, she can guess the bipartition correctly only with a chance of $\tfrac{1}{3}$ in the tripartite case, and more generally $(2^{n-1}-1)^{-1}$ in the $n$-partite scenario. Evidently, this guessing probability rapidly approaches 0 as $n$ increases, making it extremely difficult for Eve to succeed with this eavesdropping strategy for large $n$.

{\em Application II: bipartite leakage-resilient QKD.}--Next, let us describe a quantum key distribution (QKD) protocol between two parties, A and B, which is as leakage-resilient~\cite{LeakageResilience} as one could hope for. The protocol consists of: (i) preparation of many copies of $\ket{\rm GHZ_n}$, (ii) for each of these $n$-partite systems, a {\em randomly chosen subset}, say, $k$ of the $n$ subsystems are distributed to A, while the remaining $n-k$ subsystems are distributed to B, (iii) for each of these subsystems, A and B randomly measure $\sigma_x$ or $\sigma_y$,  (iv) both {\em parties} announce their sum of inputs, (v) for each $n$-partite system distributed from the source, A and B compute the product of their local measurement outcomes  and deduce, using Eq.~\eqref{Eq:GHZ:Correlations},  the shared secret bit upon learning the sum of inputs of the other party.

As with the MSS protocol described above, the secret key is established through the perfect (anti) correlation present in the product of the outputs. Moreover, the gist of the protocol only relies on the correlation given by Eq.~\eqref{Eq:GHZ:Correlations}, rather than the actual state and measurement giving rise to this correlation, rendering the protocol ideal for device-independent analysis.
However, in contrast with usual device-independent cryptography where leakage of information is not allowed,  the above protocol is as leakage-resilient as one can hope for --- the adversary Eve can certainly recover the secret key if all the output bits from either party leak to her, but if she misses merely one output bit from each party, the additional information that she gains from the leakage cannot improve her guess of the secret key. Now, if we {\em assume} that Eve has no control over how the subsystems are distributed in step (ii)\footnote{Instead of this assumption, A \& B can employ {\em additional} measurement settings, cf.~\cite{Acin:NJP:2006}, to certify that the overall correlations indeed exhibit genuine multipartite nonlocality and they are then again not susceptible to such an attack.} but otherwise only constrained by the non-signaling principle, then as with the MSS protocol, for $n$ sufficiently large, her advantage of preparing some biseparable, non-signaling boxes for A \& B is minimal.

{\em Discussion.}-- Let us now comment on some possible directions for future research. Clearly, we have only provided intuitions on why the  protocols proposed above may be secure even in a device-independent setting. For odd $n\ge3$, since the GHZ correlations violate the Mermin-Bell inequality maximally (see Appendix \ref{App:BIV}), the result of Franz \etal~\cite{Franz:PRL:2011} implies that these correlations are necessarily monogamous with respect to any potential quantum eavesdropper. This strongly suggests that if we assume an independent-and-identically-distributed (i.i.d) scenario, a formal security proof of these protocols against a quantum adversary may be given even in the case with noisy correlations,\footnote{Due to the noise robustness of the Mermin-Bell violation of $\vecP^n_{\rm GHZ}(\vec{a}|\vec{x})$, the ANL of $\vecP^n_{\rm GHZ}(\vec{a}|\vec{x})$ is also extremely robust to noise.}   and in a device-independent setting. Evidently, a security proof without this assumption is even more desirable, and a possible path towards this is to prove that the protocols are even secure against an adversary that is only constrained by the non-signaling principle~\cite{NS}. Our arguments on why the protocols are not immediately susceptible to a straightforward attack by such an eavesdropper, despite the fact that the correlations are biseparable, is an evidence pointing in this direction.

For leakage-resilient QKD, one could also imagine, instead of the above protocol,  doing an existing QKD protocol many times in parallel and then using the XOR of the secret key bits to generate the final secret key. Although such a protocol requires many more qubits to establish the final secret key, it can clearly offer high level of leakage resilience. How would such a protocol perform compared with the above protocol based on $\ket{\rm GHZ_n}$? This certainly deserves some further investigation.

Coming back to ANL itself, let us note that the requirement of (1) nonlocality and (2) biseparability with respect to all bipartitions {\em may arguably not}, by themselves, imply that an outsider cannot attribute unambiguously the nonlocality to {\em any definite subset(s)} of the $n$ parties. For instance, one may start with the tripartite GHZ correlation $P^3_{\rm GHZ}(\vec{a}|\vec{x})$, cf. Eq.~\eqref{Eq:GHZ:Correlations}, and trivially construct an example $P'=P^3_{\rm GHZ}(\vec{a}|\vec{x}) \prod_{i=4}^n P(a_i|x_i)$ for arbitrary $n$ parties by introducing parties that are uncorrelated with the first three. While such an $n$-partite correlation $P'$ indeed satisfies the two requirements stated above, one can unambiguously attribute  the nonlocality present only to the three parties that give rise to $P^3_{\rm GHZ}(\vec{a}|\vec{x})$. Note, however, that such an identification is {\em incomplete} since the production of such a biseparable correlation only requires the nonlocal collaboration between two parties, and it is still impossible for an outsider to determine which two parties have collaborated nonlocally in producing the given correlation (Fig.~\ref{Fig:Polytopes} and Fig.~\ref{Fig:ANL}). A more precise definition of ANL may thus require also a specification of the extent (size) of the nonlocal resource needed in producing the given correlation, a task that shall be pursued elsewhere~\cite{Curchod:IP}. For our GHZ examples, except for the cases where  $n$ is even with $\frac{n}{2}$ odd, it can be shown  (see Appendix~\ref{App:m-sep}) using the result of Ref.~\cite{Bancal:PRL:2009}  that the correlations of Eq.~\eqref{Eq:GHZ:Correlations} are not triseparable, i.e., not producible by a partitioning of the parties into three groups (where only parties within the same group are allowed to collaborate nonlocally). Hence, the generation of these correlations indeed requires the nonlocal collaboration of at least $\lceil \frac{n}{2} \rceil$ parties in one group; an analogous statement for the remaining cases would be desirable.

\begin{acknowledgments}
We are grateful to David Jennings for suggesting the terminology ``Anonymous Nonlocality", and to Rotem Arnon-Friedman, Jean-Daniel Bancal, Nicolas Brunner, Stefano Pironio as well as Renato Renner   for stimulating discussions. This work is supported by the Swiss NCCR ``Quantum Science and Technology", and the CHIST-ERA DIQIP. FJC acknowledges support from the John Templeton Foundation. YCL and FJC contribute equally towards this work.
\end{acknowledgments}

\appendix

\section{An explicit biseparable decomposition of the $n$-partite GHZ correlations}
\label{App:NS-Bisep}

For the $n$-partite GHZ state  and the situation where all parties measure either the 0th-observable $\sigma_x$ or the 1st observable $\sigma_y$, the resulting correlation of Eq.~\eqref{Eq:GHZ:Correlations} can be rewritten in terms of the {\em correlator}, i.e., the expectation value of the product of outcomes:\footnote{To arrive at this $n$-partite correlator, see, eg., Eq. (23) of~\cite{Wallman:PRA:2011}.}
\begin{equation}
	\!\!E(\vec{x})=\sum_{a_1',a_2',\ldots,a_n'=0,1} (-1)^{\sum_i a_i'} P(\vec{a}'|\vec{x})
	=\cos\left(\bfx\frac{\pi}{2}\right)\tag{\ref{Eq:GHZ:Correlations}}
\end{equation}
where for conciseness of subsequent presentation we have used, instead, $a_i'=\frac{a_i+1}{2}=0,1$ to denote the output and as before,  $\bfx=\sum_i x_i$ to  denote the sum of inputs. Note that all the full $n$-partite correlators depend only on the parity of $\bfx$ and $\bfx/2$ whereas all the marginal correlators vanish. 

Here we give a proof that the above correlation is biseparable with respect to {\em all} bipartitions whenever parties in each group are allowed to share arbitrary post-quantum but non-signaling (NS) resources, while parties in different groups can only be correlated through shared randomness. Note that the biseparability of Eq.~\eqref{Eq:GHZ:Correlations} under the NS constraint implies that if parties in the same group are allowed to share a stronger resource, such as a Svetlichny resource~\cite{Svetlichny}, or some other one-way signaling resource discussed in Refs.~\cite{Gallego:PRL:070401,Bancal:PRA:014102}, the correlation must remain biseparable.

Let us define the four families of $n$-partite NS boxes, labeled by $\mu_1$, $\mu_2$, $\mu_3$ and $\mu_4$:
\begin{equation}\label{Eq:n-PRBoxes}
\begin{split}
	P^n_{\mu_1}(\vec{a}'|\vec{x})&=\frac{1}{2^{n-1}}\delta_{\sum_{i=1}^n {a_i'}-H^n_0(\vec{x})-H^n_3(\vec{x})\,{\rm mod}\,2},\\
	P^n_{\mu_2}(\vec{a}'|\vec{x})&=\frac{1}{2^{n-1}}\delta_{\sum_{i=1}^n {a_i'}-H^n_0(\vec{x})-H^n_1(\vec{x})\,{\rm mod}\,2},\\
	P^n_{\mu_3}(\vec{a}'|\vec{x})&=\frac{1}{2^{n-1}}\delta_{\sum_{i=1}^n {a_i'}-H^n_1(\vec{x})-H^n_2(\vec{x})\,{\rm mod}\,2},\\
	P^n_{\mu_4}(\vec{a}'|\vec{x})&=\frac{1}{2^{n-1}}\delta_{\sum_{i=1}^n {a_i'}-H^n_2(\vec{x})-H^n_3(\vec{x})\,{\rm mod}\,2},
\end{split}
\end{equation}
where $H^n_\ell(\vec{x})=\sum_{j=0}^{\lfloor \frac{n-\ell}{4}\rfloor} F(4j+\ell,\vec{x})$,
\begin{equation}
	F(k,\vec{x})=\sum_{G} \prod_{i\in G} x_i \prod_{j\in G'} (x_j+1)
\end{equation}
and the sum $\sum_{G}$ is over all $G\subseteq [n]=\{1,2,\ldots,n\}$ with group size $|G|=k$, and $G'$ is the complement of $G$ in $[n]$. Essentially, each term involved in the summand in $F(k,\vec{x})$, and hence $H^n_\ell(\vec{x})$ defines a distinct combination of inputs $\vec{x}=\vec{x}'$ such that $H^n_\ell(\vec{x}')=1\,{\rm mod}\,2$, and hence making the outputs anti-correlated. For instance, $F(0,\vec{x})$ only makes a nontrivial combination to $H^n_0(\vec{x})$ if all the inputs $x_i$ are 0.

From Eq.~\eqref{Eq:n-PRBoxes}, it is easy to verify that for all $1\le k\le n-1$, the $k$-partite marginals of $P^n_j(\vec{a}'|\vec{x})$ are $1/2^k$ and these correlations indeed define NS probability distributions. Moreover, from Eq.~\eqref{Eq:n-PRBoxes} and these marginal distributions, one can show that these NS boxes give rise to vanishing marginal correlators and the following full $n$-partite correlators:
\begin{equation}\label{Eq:NSBox:Correlator}
\begin{split}
	E(\vec{x})_{\mu_1}=(-1)^{H^n_0(\vec{x})\oplus H^n_3(\vec{x})}=-E(\vec{x})_{\mu_3},\\
	E(\vec{x})_{\mu_2}=(-1)^{H^n_0(\vec{x})\oplus H^n_1(\vec{x})}=-E(\vec{x})_{\mu_4},
\end{split}	
\end{equation}
where in Eq.~\eqref{Eq:NSBox:Correlator},  $\oplus$ denotes sum modulo 2 and in arriving at the second equality in each line, we have employed the  identity $\sum_{j=0}^n F(j) =1$ that holds for all $n$-bit strings $\vec{x}$.\footnote{This last sum involves all possible combinations of inputs and thus for all input bit strings $\vec{x}$, there is exactly one term in the expression that does not vanish, therefore giving the identity.} 
To gain some intuition on these NS boxes, we note that for $n=1$, the $\mu_{1/3}$ boxes correspond to the deterministic strategies $a'=x\oplus1$ and $a'=x$ whereas the $\mu_{2/4}$ boxes correspond to the deterministic strategies $a'=1$ and $a'=0$. Similarly, for $n=2$,  the $\mu_{1/3}$ boxes correspond to the PR boxes defined by $a'_1+a'_2=(x_1+1)(x_2+1)$ and $a'_1+a'_2=(x_1+1)(x_2+1)\oplus1$ whereas the $\mu_{2/4}$ boxes correspond to the PR boxes defined by $a'_1+a'_2=x_1x_2\oplus1$ and $a'_1+a'_2=x_1x_2$. For $n=3$, all these NS boxes correspond to some version of NS box 46 described in Ref.~\cite{Pironio:JPA:2011}. It is conceivable that these boxes are extremal NS distributions for all $n$.

To reproduce the correlations given in Eq.~\eqref{Eq:GHZ:Correlations} using biseparable $\NS$ resources with $k$ parties in one group and the remaining $(n-k)$ parties in the other group, it suffices to consider an equal-weight mixture of the following four strategies:
\begin{enumerate}
	\item The group of $k$ parties share the $k$-partite version of the $\mu_1$ box and the remaining parties share the $(n-k)$-partite version of the $\mu_2$ box.
	\item The group of $k$ parties share the $k$-partite version of the $\mu_3$ box and the remaining parties share the $(n-k)$-partite version of the $\mu_4$ box.
	\item The group of $k$ parties share the $k$-partite version of the $\mu_2$ box and the remaining parties share the $(n-k)$-partite version of the $\mu_1$ box.
	\item The group of $k$ parties share the $k$-partite version of the $\mu_4$ box and the remaining parties share the $(n-k)$-partite version of the $\mu_3$ box.
\end{enumerate}
For $n=3$, the above strategy corresponds to a mixture of 4 different versions of the NS box 2 in Ref.~\cite{Pironio:JPA:2011}. In general, to verify that the above strategy indeed gives rise to Eq.~\eqref{Eq:GHZ:Correlations}, we first remark that each of these strategies also reproduces Eq.~\eqref{Eq:GHZ:Correlations} for the case when $\sum_i x_i$ is even. To see this, we use the fact that  NS box $\mu_1$ gives anti-correlation (i.e., expectation value -1) only if either  $\sum_i x_i/2$ or $ (1+\sum_i x_i)/2$ is even; NS box $\mu_2$ gives anti-correlation only if $\sum_i x_i/2$ is even or $(1+\sum_i x_i)/2$ is odd; NS box $\mu_3$ gives anti-correlation only if either $\sum_i x_i/2$  or $(1+\sum_i x_i)/2$ is odd; NS box $\mu_4$ gives anti-correlation  only if $\sum_i x_i/2$ is odd or $(1+\sum_i x_i)/2$ is even. Moreover, since strategy 1 and 3 are such that the correlation produced by parties in the same group are exactly opposite (likewise for strategy 2 and 4), we see that all the less-than-$n$-partite  correlators, as well as the full $n$-partite correlator when $\sum_{i=1}^n x_i$ is odd, indeed vanishes as claimed.

\section{Mermin-Bell violation of the GHZ correlations}
\label{App:BIV}

Here, we compute the quantum expectation value of the GHZ correlations for the Mermin Bell inequality~\cite{MABK1,MABK2} (here written in the form derived in~\cite{Wallman:PRA:2011})\footnote{$\B^n_{+}$  is the same Bell expression as the usual one obtained through the recursive formula~\cite{MABK2}; it can also be obtained by flipping all the inputs in $\B^n_{-}$.}
\begin{equation}\label{Ineq:MABK}
	 \left|\B^n_{\pm}\right|=2^{\frac{1-n}{2}}\Bigl|\sum_{\vec{x} \in\{0,1\}^n}\!\!\!\!
	 \cos\left\{\frac{\pi}{4}\left[1\pm(n-2\bfx)\right]\right\}\!E\left(\vec{x}\right)\Bigr| \le1.
\end{equation}
The above Bell expression can be rewritten as:
\begin{align*}
	 2^{\frac{n-1}{2}}\left|\B^n_{\pm}\right|=&\Bigl|\sum_{\vec{x} \in\{0,1\}^n}\!\!\!\!
	 \cos\left\{\frac{\pi}{4}\left[1\pm(n-2\bfx)\right]\right\}\!E(\vec{x})\Bigr|,\\
	 =&\Bigl|\sum_{\vec{x} \in\{0,1\}^n}\!\!\!\!
	 \cos\left[\frac{\pi}{4}(1\pm n)\right]\cos\left(\bfx\frac{\pi}{2}\right)\!E(\vec{x}) \\
	 &\pm\sum_{\vec{x} \in\{0,1\}^n}\!\!\!\!
	 \sin\left[\frac{\pi}{4}(1\pm n)\right]\sin\left(\bfx\frac{\pi}{2}\right)\!E(\vec{x})\Bigr|. 
\end{align*}
For the GHZ correlation of Eq.~\eqref{Eq:GHZ:Correlations}, this simplifies to
\begin{align*}
	 \left|\B^n_{\pm}\right|
	 =&2^{\frac{1-n}{2}}\Bigl|\sum_{\vec{x} \in\{0,1\}^n,\, \bfx\,\text{even}}\!\!\!\!
	 \cos\left[\frac{\pi}{4}(1\pm n)\right]\cos^2\left(\bfx\frac{\pi}{2}\right)\Bigr|,\\
	 =&2^{\frac{n-1}{2}}\Bigl|\cos\left[\frac{\pi}{4}(1\pm n)\right]\Bigr|,
\end{align*}
giving 
\begin{equation}\label{Eq:MerminViolation}
\max_{\pm} \left|\B^n_\pm\right|=\left\{\begin{array}{r@{\quad \quad}}
2^{\frac{n-1}{2}}:
 n\,\rm{odd} \\
2^{\frac{n-2}{2}}: n~\rm{even} \\  \end{array}\right. ,
\end{equation}
i.e., achieving maximal~\cite{Werner:PRA:2000} possible quantum value of $\left|\B^n_\pm\right|$ for odd $n$.

\section{Quantum biseparable bound of the $n$-partite Mermin-Bell expression}
\label{App:Quantum:BS}

For arbitrary odd $n\ge3$, the Mermin-Bell expression $\B^n_+$ given on the left-hand-side of Eq.~\eqref{Ineq:MABK} is equivalent to a special case of a general family of permutationally invariant Bell expression described in Eq.~(22) of~\cite{Bancal:JPA:2012},
\begin{equation}
	\Omega_{n,2,2;\delta_{\bfx,0}\cdot r}=2^{n-2}-2^{\frac{n-3}{2}}\B^n_+
\end{equation}

From Eq.~(23) of Ref.~\cite{Bancal:JPA:2012}, it can be shown that the above expression admits the following upper bound on the quantum biseparable bound:
\begin{equation}
	\Omega_{n,2,2;\delta_{\bfx,0}\cdot r}\ge 2^{n-3}(2-\sqrt{2}).
\end{equation}
Combining these two equations and after some straightforward computations, we get the following upper bound on the quantum biseparable bound for the Mermin-Bell expression:
\begin{equation}\label{Ineq:Mermin:DIEW}
	\B^n_+\le 2^{\frac{n}{2}-1}.
\end{equation}

For arbitrary even $n\ge2$, the Mermin-Bell expression $\B^n_+$ given on the left-hand-side of Eq.~\eqref{Ineq:MABK} is equivalent to the following Bell expression described in Eq.~(1) of Ref.~\cite{Bancal:JPA:2012},
\begin{equation}
	\mathcal{I}_{n,2,2}=2^{n-1}-2^{\frac{n-2}{2}}\B^n_+
\end{equation}

From Eq.~(25) of Ref.~\cite{Bancal:JPA:2012}, we know that the above expression admits the following upper bound on the quantum biseparable bound:
\begin{equation}
	\mathcal{I}_{n,2,2}\ge 2^{n-2}.
\end{equation}
Combining these two equations, we arrive, again, at Eq.~\eqref{Ineq:Mermin:DIEW}.

To see that the biseparable bound of Eq.~\eqref{Ineq:Mermin:DIEW} is tight, it suffices to note that the biseparable quantum state
\begin{equation}
	\ket{\psi}=\ket{\mbox{GHZ}_{n-1}}\otimes\ket{0}
\end{equation}
and the local observables
\begin{equation}
\begin{split}
	A_{x_i}&=\cos \alpha_{x_i} \sigma_x +\sin\alpha_{x_i}\sigma_y\quad\forall\,\, i=1,\ldots,n-1,\\
	A_{x_i}&=\beta_{x_i}\id\quad\mbox{for}\quad i=n.
\end{split}	
\end{equation}
with $\alpha_0=-\frac{\pi}{4(n-1)}$, $\alpha_1=-\frac{\pi}{2}-\frac{\pi}{4(n-1)}$, $\beta_0=-\sqrt{2}\sin\frac{n\pi}{4}$, and $\beta_1=\sqrt{2}\cos\frac{n\pi}{4}$ indeed give rise to a quantum value of $\B^n_+$ of $2^{\frac{n}{2}-1}$. Since $\B^n_-$ can be obtained from $\B^n_+$ by flipping all the inputs, the same quantum biseparable bound holds for $\B^n_-$.

Since the GHZ correlations of Eq.~\eqref{Eq:GHZ:Correlations} give Eq.~\eqref{Eq:MerminViolation},  we see that for odd $n$, the generation of these correlations necessarily requires a genuinely $n$-partite entangled state, independent of the underlying Hilbert space dimension.

\section{$m$-separability and multipartite nonlocality underlying the $n$-partite GHZ correlations}
\label{App:m-sep}

For odd $n$, we know from the main theorem of~\cite{Bancal:PRL:2009} that a quantum violation of $\left|\B^n_{\pm}\right|=2^{\frac{n-1}{2}}$ implies that it is impossible to reproduce these GHZ correlations using any 3-separable resource (i.e., a partitioning of the parties into three groups, and where the parties within each group can share even arbitrary  nonlocal resource).

For even $n$, let us  evaluate the the quantum value of the following Bell expression~\cite{Bancal:PRL:2009}:
\begin{align}
	\left|\B^n_{\Sigma}\right|=&\frac{1}{\sqrt{2}} \left|\B^n_+ + \B^n_-\right|,\nonumber\\
	=&\frac{1}{\sqrt{2}}\Bigl|\sum_{\vec{x} \in\{0,1\}^n}\sum_{s=0,1} \cos\frac{\pi}{4}\!\left[1+(-1)^s(n-2\bfx)\right]E(\vec{x})\Bigr|,\nonumber\\
	=&\frac{1}{\sqrt{2}}\Bigl|\sum_{\vec{x} \in\{0,1\}^n}\!\!\!\!2
	 \cos\frac{\pi}{4}\cos\left[\frac{\pi}{4}(n-2\bfx)\right]\!E(\vec{x})\Bigr|,\nonumber
\end{align}
\begin{align}
	=&\Bigl|\sum_{\vec{x} \in\{0,1\}^n}\!\!\!\!
	 \cos\left[\frac{\pi}{4}(n-2\bfx)\right]\!E(\vec{x})\Bigr|.
\end{align}
For even $n$  and $E(\vec{x})$ of Eq.~\eqref{Eq:GHZ:Correlations}, this becomes
\begin{equation*}
\begin{split}
	&\Bigl|\sum_{\vec{x} \in\{0,1\}^n,\, \bfx\,\text{even}}\!\!\!\!
	 \cos\frac{n\pi}{4}\cos^2 \bfx\frac{\pi}{2}\Bigr|=2^{n-1}\left| \cos\frac{n\pi}{4}\right|, 
\end{split}
\end{equation*}
giving a value of $2^{n-1}$ for even $\frac{n}{2}$ and 0 for odd $\frac{n}{2}$.

Again, note from the main theorem of Ref.~\cite{Bancal:PRL:2009} that for even $n$, any correlation producible by a partition of the $n$ parties into 3 groups (each sharing some Svetlichny resource $\S$~\cite{Svetlichny,Gallego:PRL:070401,Bancal:PRA:014102}) can at most give a value of $\B^n_{\Sigma}=2^{n-2}$. This means that, as with odd $n$, the $n$-partite GHZ correlation for even $n$ with even $\frac{n}{2}$ is not producible by any partition of the parties into 3 groups, even if parties in each group are allowed to share whatever nonlocal resource.

Together with the biseparable decomposition obtained for these correlations, the above results on $m$-separability imply that  for (1) odd $n$ and (2) even $n$ with even $\frac{n}{2}$, generation of the GHZ correlations of Eq.~\eqref{Eq:GHZ:Correlations} requires the nonlocal collaboration of at least  $\lceil\frac{n}{2}\rceil$ parties in one group.

\end{document}